# Causal Analysis of First-Year Course Approval Delays in an Engineering Major Through Inference Techniques.

Hugo Roger Paz

Professor and researcher at the Faculty of Exact Sciences and Technology of the National University of Tucumán.
Doctoral student in Exact Sciences and Engineering from FACET-UNT.
Email: hpaz@herrera.unt.edu.ar
ORCID: https://orcid.org/0000-0003-1237-7983

## Abstract

The study addresses the problem of delays in the approval of first-year courses in the Civil Engineering Major at the National University of Tucumán, Argentina. Students take an average of 5 years to pass these subjects. Using the DoWhy and Causal Discovery Toolbox tools, we looked to identify the underlying causes of these delays. The analysis revealed that the regulatory structure of the program and the evaluation methods play a crucial role in this delay. Specifically, the accumulation of regular subjects without passing a final exam was identified as a key factor. These findings can guide interventions to improve student success rates and the effectiveness of the education system in general.

## 1. Introduction

This article delves into the complexities of delayed curricular progression in higher education by investigating the underlying causes in the context of the 2005 updated curriculum of the Civil Engineering program at the National University of Tucumán. The five-and-a-half-year study plan, divided into eleven semesters, is governed by a Correlative System that sets up prerequisites for enrolling in courses. These prerequisites can be "regularized", showing the approval of partial evaluations or practical tasks, or "approved", showing the successful completion of final exams which implies that you have passed the course.

In this study, we start from the hypothesis that the existing evaluation system contributes to the accumulation of "regular" subjects, but not passed among students. We keep that this phenomenon arises from the lack of time available for adequate preparation for final exams, which, in turn, causes delays in the course approval process. Our main goal is to empirically demonstrate this underlying hypothesis, through the application of rigorous techniques of causal inference. Through this approach, we seek to shed light on the causal connections between the evaluation system, the accumulation of pending subjects and the delays in curricular progression.

As highlighted Pearl (2009), causal inference allows the identification of causal relationships beyond mere correlations. The emphasis on causal inference arises from the need to discover the true causal factors that drive delays in curriculum progression. "Observational research findings may be inconsistent or consistent but are unlikely to reflect true cause-and-effect relationships"(Hammerton and Munafo, 2021).

As Oloriz et al conclude in a paper where they study the relationship between the academic performance of entrants in engineering Majors and the dropout of university studies, "there is an important correlation between academic performance in the first quarter... and the dropout that occurs during the second, third and fourth quarters"(Oloriz et al., 2007, p. 1), that is why it is necessary to work on the reduction of academic failure since it leads, inexorably, to the abandonment of higher education.

## 2.- Method

### 2.1.- Population and Sample

The data under analysis come from the academic trajectories of 1,615 students enrolled in the Civil Engineering major, covering a diverse spectrum of Major advancements, from first-year students to those nearing graduation, as well as individuals who have dropped their training and those who have completed their educational process. These data were obtained through the Student and Administrative Management System (SIGEA), which collects and centralizes relevant information on the academic progress and management of students in the program.

Within the framework of this research, the academic performance of university students has been examined over a period of fourteen years, from 2004 to 2019. To ensure the accuracy of the results, an exclusion was made of those students who had obtained the approval of subjects through an equivalency system, due to modifications in their study plan, academic trajectory or educational institution. This measure was implemented considering that these students could have entered the program with subjects already passed, which could affect their progress and performance compared to their peers. For the analysis, we tried to select a set of 1,343 academic histories out of a total of 1,615.

In the present context, those students who have completed their studies within the period under evaluation (2005-2019) are classified as "graduates". For their part, students who have kept their enrollment in the academic program until 2019 and have not yet completed their training are considered "permanent".

Ultimately, the situation of "dropout" is defined for those students who do not register any enrollment in the year 2020 and have not completed their educational process. In adopting this approach, we follow the proposal of Gonzalez Fiegehen (2007)for students who enrolled in 2019, representing only 5% of the total analyzed(Peace, 2022). For the rest of the students, a less rigorous criterion is used, including those who resume their enrollment in the degree and resume their studies after a period of inactivity. The calculation of the time lapses has been based on the difference between the last activity documented in the SIGEA system and the date of admission to the academic program.

### 2.2.- Analysis technique

In the world of data analysis, understanding causal relationships is essential to making informed decisions and understanding the effects of various actions. In this context, Dowhy stands as a tool that uses solid theoretical foundations to conduct advanced causal analysis, unraveling hidden connections in data and providing deep insight into cause-and-effect relationships.

Dowhy is based on the theory of design of experiments and causal inference. His approach is based on the concept of "intervention", where variables are manipulated to see how they affect other variables in a system.

Once a causal effect has been found, different estimation methods can be adopted compatible with the identification strategy. To estimate the average causal effect, DoWhy supports the following methods: Distance-based matching(Chen et al., 2007), Propensity-based Stratification(Desai et al., 2017), Propensity Score Matching(Caliendo & Kopeinig, 2008; Rosenbaum & Rubin, 1983), Linear Regression, Generalized Linear Models (including logistic regression) (Chernozhukov et al., 2022), Binary Instrument/Wald Estimator(Jiang et al., 2023), Regression discontinuity(Imbens & Lemieux, 2008), Two-stage linear regression(Angrist & Imbens, 1995).

Dowhy performs a wide range of causal inference techniques, including estimating the Average Treatment Effect (ATE).(Crump et al., 2008)and other measures of causal impact. The tool also allows the estimation of confidence intervals and tests of statistical significance to support the results obtained. The sequence of analysis entails the execution of the following steps:

**A. Definition of the Causal Graph:** The first step is to model the causal graph, which is the relationships between the variables of interest and how they influence each other. This helps to visualize the system and understand the underlying connections. A causal graph models the causal relationships, or "cause-effect relationships" present in a system or problem domain. In DoWhy, the causal graph is required to be a directed acyclic graph (DAG) where an edge X→Y implies that X causes Y. Statistically, a causal graph encodes conditional independence relationships between variables.

Figure 1. Structure of the Causal Model

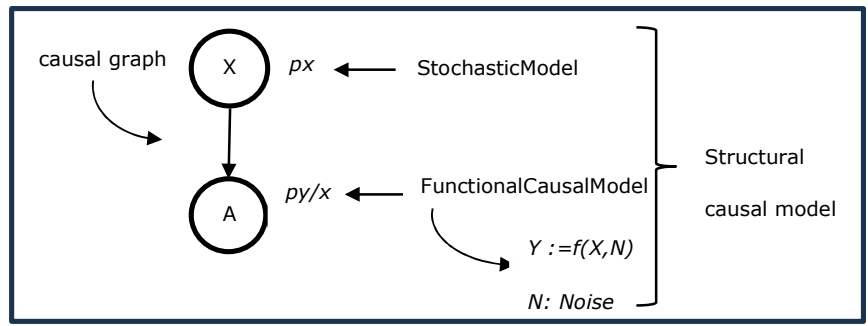

Source: DoWhy User Guide. Available at:
https://www.pywhy.org/dowhy/v0.10/user_guide/modeling_gcm/index.html

One of the essential elements in the causal analysis process with Dowhy is the definition of the causal graph, which captures the interactions and relationships between the variables of interest. However, this step is not an isolated task; rather, it requires deep prior knowledge about the problem at hand and possible causal relationships, based on experience and intuition.

The process of defining a causal graph within Dowhy is like creating a map that will guide the analysis towards understanding the underlying causal relationships. This map cannot be generated solely by algorithms or automated approaches, as it requires a deep understanding of the specific domain. This is where prior knowledge comes into play.

Foreknowledge is the amalgamation of experience, intuition, and informed assumptions. Those who are familiar with the problem at hand can find the likely influencing variables and likely directions of causal relationships. Some assumptions may be based on earlier research, fundamental theories, or even the simplest observations.

The definition of the causal graph is based on the idea that many relationships in the real world can be interpreted as intervenable causality (IC) graphs. An IC chart is a visual representation of the causal relationships between variables, with arrows showing direct influences. In this context, modeling the causal graph in Dowhy involves mapping the cause-and-effect relationships of the problem in a graphical format.

The graph definition process is not merely technical; it is an art backed by experience. Domain experts are those who can bring valuable ideas and insight to the task. Its ability to find non-obvious connections and capture the subtleties of causal interactions enriches the analysis and ensures that the graph is a faithful representation of reality.

The definition of the causal graph in Dowhy is a fundamental part of the process of causal analysis. It requires input from prior knowledge based on experience and informed assumptions. This collaborative approach between human expertise and technological tools ensures that causal analysis is a rigorous and results-oriented process in search of meaningful connections between variables.

**B. Identification and Estimation of the Causal Effect:** Once the graph is defined, Dowhy uses observational data and causal inference methods to estimate the effect of the interventions on the variables of interest. This involves the selection of variables, the calculation of propensity scores if necessary, and the application of adjustment techniques.

Once the causal graph is defined, the treatment (independent variable) and the outcome (dependent variable) are selected to be investigated to estimate its causal effect. Dowhy is based on the concept of intervention, where the manipulation of the treatment is simulated to see its impact on the result.

In many cases, observational data may be subject to bias due to self-selection in treatments. Dowhy uses the "propensity score matching" method to estimate the probability of receiving the treatment (propensity score) based on other observed variables. This helps create comparable groups and adjust for potential biases.

Dowhy offers a variety of causal effects estimation methods, such as the "Matching" method, "Regression Discontinuity", "Instrumental Variables", among others. These methods are applied to adjusted and controlled observational data, to assess how the outcome changes when the treatment is manipulated.

After estimating the causal effect, Dowhy allows you to assess the statistical significance of the results. This is carried out by estimating confidence intervals and performing hypothesis tests to determine if the observed effect is statistically significant.

Once the results are obtained, Dowhy facilitates the interpretation of the estimated causal effects. The obtained values, together with confidence intervals and significance tests, provide valuable information about how treatment affects outcome and whether this relationship is plausible from a causal perspective.

Dowhy performs a comprehensive process of causal effect identification and estimation, combining causal theory with advanced statistical methods. Through the definition of the causal graph, the selection of variables, the use of the propensity score and the application of estimation methods, Dowhy sheds light on causal relationships in observational data.

**C. Evaluation of the Validity of the Model:** Dowhy allows you to assess the statistical significance of the results obtained, which provides an objective measure of the reliability of the estimated causal effects.

It should be noted that the causal part does not come from data. It comes from the assumptions leading to the identification made at point A. The data is used simply for statistical estimation. Therefore, it becomes essential to check whether our assumptions were correct in the first step or not. What happens when there is another common cause? What happens when the treatment itself was placebo? To do this, three methods of refuting results are used:

**Method 1: Random common cause.** Add randomly drawn covariates to the data and rerun the analysis to see whether or not the causal estimate changes. If our assumption was originally correct, then the causal estimate shouldn't change much.

**Method 2: Refuting placebo treatment.** randomly assigns any covariate as treatment and reruns the analysis. If our assumptions were correct, then this newfound estimate should go to 0.

**Method 3: Data subset refuter.** creates subsets of data (similar to cross-validation) and checks whether causal estimates vary between subsets. If our assumptions were correct, there shouldn't be much variation.

The application of the aforementioned techniques was carried out through the development of a code in Python language(Van Rossum & Drake Jr, 1995). The specific DoWhy libraries were used for the development of the code.(Sharma & Kiciman, 2020)and Causal Discovery Toolbox(Kalainathan et al., 2020).

## 3.- Results

### 3.1.- Definition of the variables of the problem

In this study, we start from the hypothesis that the existing evaluation system contributes to the accumulation of "regular" subjects, but not passed among students, which ultimately produces the causal effect of generating delays in the course approval process. This analysis will focus in this work on the delay that students suffer in the approval of the first-year courses.

From this point of view, the variables "treatment: T" (Accumulation of Regular Subjects) and the variable "result: Y" (Time it takes students to pass the subjects corresponding to the first year of the degree) can then be defined. The T $\rightarrow$ Y causal relationship involves other variables or "confounders" which intervene in said causal relationship, affecting both the treatment and the outcome, according to the following graph:

Figure 2. Flow rate graph (DAG)

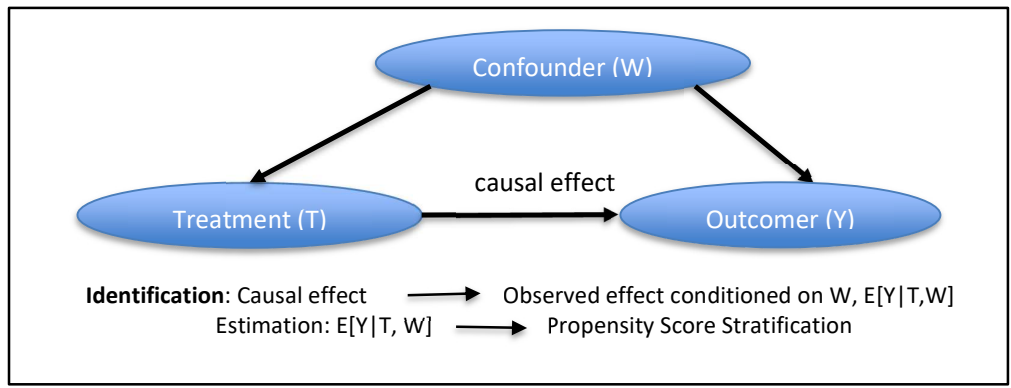

Source: DoWhy User Guide. Available at:
https://www.pywhy.org/dowhy/v0.10/user_guide/modeling_gcm/index.html

Based on this, the questions that arise are the following:

What is the impact that the accumulation of regular subjects has on the time it takes for students to pass all the subjects of the first year of the degree?

And the equivalent counterfactual question is:

If students do not have less than the average amount of accumulation of regular subjects, what is the probability that the approval time will decrease? Or vice versa. In formal language, we are interested in the average effect of treatment on students (ATE).

For the analysis, the academic histories were used (see section 2.1.), from which the following data were extracted for 1343 students:

> Cohort
> Gender
> Age
> Time in Major
> Approved Activities
> Average Notes
> Total Number Coursed
> Number Re Coursed
> Regular Number
> Free Number
> Total appealed over total completed.
> Total Regular-Total Coursed

➢ Number of Exams
➢ Number of promotions
➢ Number Approved
➢ number of failures
➢ Number of Absentees
➢ Number of Promotions Out of Total Promotional Subjects
➢ Number of Promotions Out of Total Subjects
➢ Passed Exam Out of Total Exam
➢ Exam Promoted Over Total Exam
➢ Failed Exam Over Total Exam
➢ Absent Exam Over Total Exam
➢ has promotions.
➢ Has Free More 10
➢ Has Failed More 3
➢ Has Failed Minor 3
➢ Approval Time Modules 1-2 (M1 and 2 = First year of the Degree)
➢ Approval Time M1 2 More than 2 years (M1 and 2 = First year of the Degree)
➢ Maximum Accumulated Regular Subjects (MaxRegAcum)
➢ Accumulated Regular Subjects Greater than 6 (MaxRegAcumMayor6)

Regarding the treatment variables (Accumulated Regular Subjects) and result (Approval Time Modules 1 and 2), a binary distribution was made to allow the execution of the analysis. Said distribution (Accumulated Regular Subjects Greater than 6 for the treatment and Approval Time M1 2 Greater than 2 years for the result) was made based on the average value of said variables for the total sample. In this way, it is sought that the cases for both samples are balanced, since, otherwise, the implementation of the method throws inference errors (See Figures 3 and 4).

Figure 3. Histogram of Maximum Accumulated Regularities

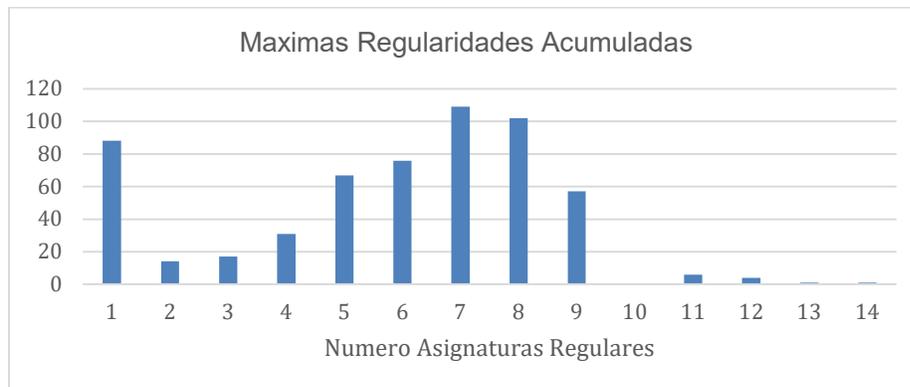

Figure 4. Histogram Approval Time Subjects Year 1

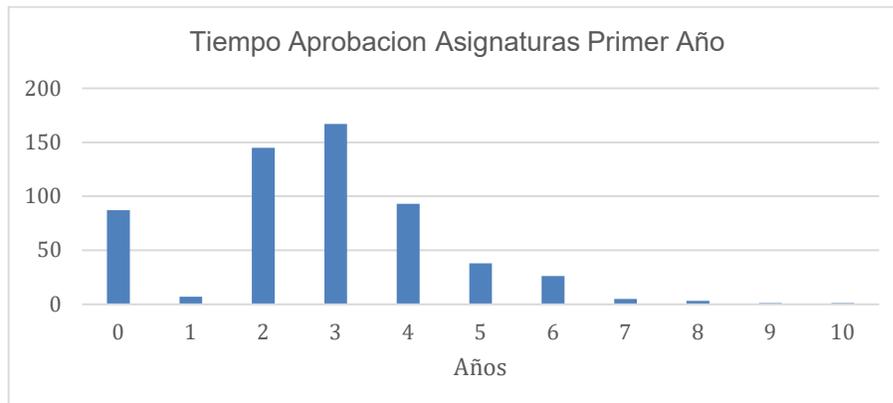

### 3.2.- Definition of the Causal Graph

The generation of the causal graph in this study has been guided by a number of fundamental assumptions about the structure of the curriculum and the interactions between assessment activities, such as course regularization and approval. In this process, treatment and outcome variables have been incorporated following the binary distribution detailed in the earlier section. These essential variables were incorporated to model the interventions and the effects saw in the study context.

It should be noted that the causal graph presented here is a first approach to causality analysis. The relationships and connections between the variables have been delineated according to the available assumptions and expert knowledge. However, to confirm the validity and robustness of these causal relationships, further refutation analyzes will be performed to allow the causal hypotheses to be rigorously evaluated.

In addition, as part of the modeling process, all the variables that were considered relevant as possible "confounders" have been introduced. The inclusion of these added variables is intended to improve the accuracy of causal inference by controlling for potential sources of bias and confounding. This adjustment strategy looks to ensure that the estimated effects are more accurately attributable to the treatment under study.

In summary, the process of generating the causal graph has involved making fundamental assumptions about the structure of the curriculum and the interactions between assessment activities. As the analysis progresses, refutation methods will be applied to confirm and strengthen the causal relationships proposed in this first graph. The inclusion of treatment, outcome and "confounders" variables looks to offer a robust and detailed representation of the underlying causal relationships in this educational context. The resulting graph can be seen in the following figure.

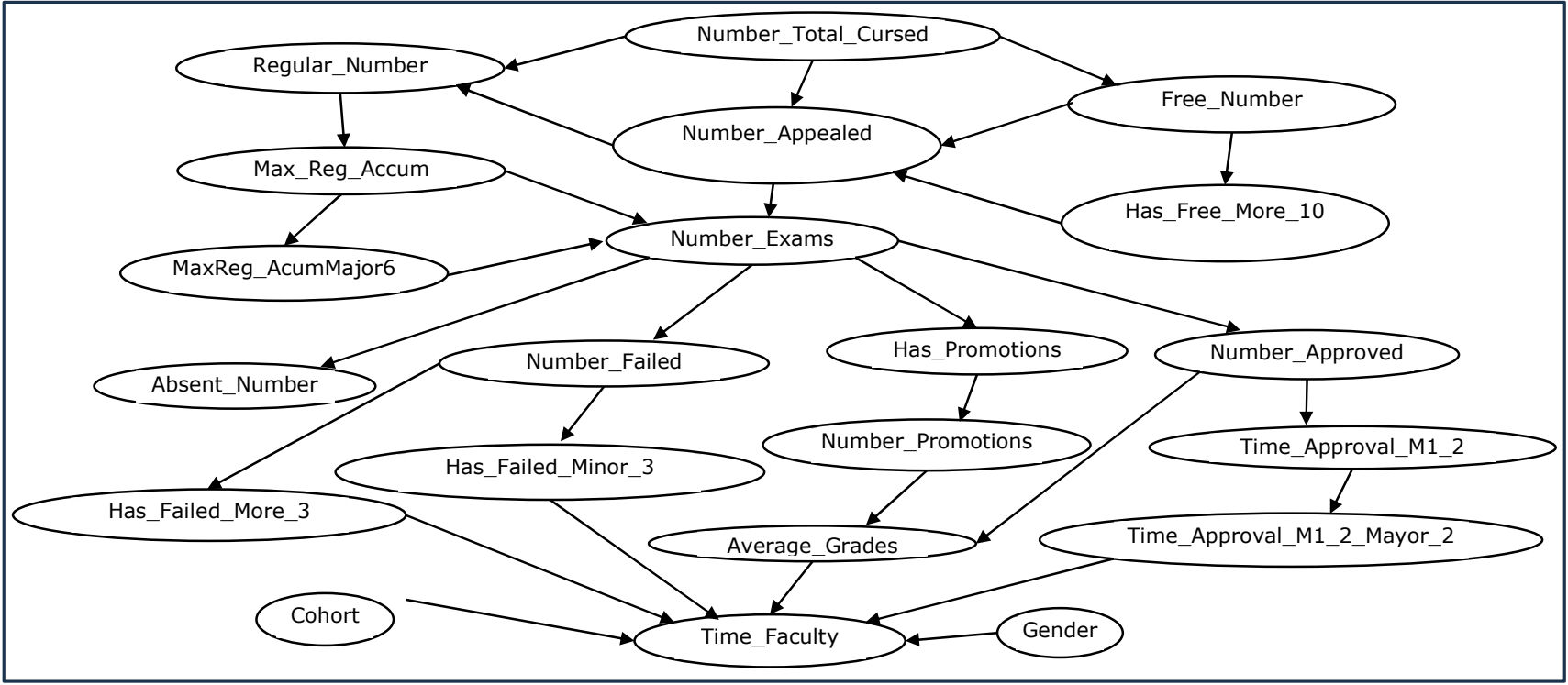

Figure 5. Direct Acyclic Causal Graph (DAG)

Source: self made.

### 3.2.- Identification and Estimation of the Causal Effect

The implementation of the Identification and Estimation of the Causal Effect was conducted by adopting specific methodological approaches, in line with the premises of scientific rigor and causal validity. In this study, two estimators, namely "backdoor.propensity_score_matching" and "backdoor.distance_matching", were employed in order to discern and quantify the causal effects under analysis. These estimators are based on the notions of "backdoor" and "matching" to address possible alternative paths of confusion and bias in observational data.(Caliendo & Kopeinig, 2008; Rosenbaum & Rubin, 1983).

Within the framework of this methodology, the concept of "backdoor" refers to variables that open additional pathways between treatment and outcome, which could lead to confusion in the estimation of the causal effect.(Luo et al., 2023). To counteract this effect, the "propensity score matching" and "distance matching" estimators were applied. The first, based on the propensity score, seeks to achieve balance between the treatment and control groups in terms of the relevant observable characteristics. The second, using the distance between the covariates, performs a more precise and calibrated matching to achieve comparability of the groups.(Caliendo & Kopeinig, 2008; Rosenbaum & Rubin, 1983).

The choice of these estimators is justified by their ability to mitigate potential bias and possible confounding variables in observational data. The theory underlying these estimators is based on the idea of adequately controlling the variables that could intervene in the causal relationship between treatment and outcome, so that the causal effect can be isolated and estimated more accurately.

In summary, the implementation of Causal Effect Identification and Estimation in this study is based on the selection of specific estimators, "backdoor.propensity_score_matching" and "backdoor.distance_matching", which are supported by causal and methodological theory. These estimators address potential confounding pathways by rigorously controlling variables using the matching technique. The application of these methodological approaches advances towards a more solid and informed understanding of the causal effects in the studied context. The results can be seen below.

**`propensity_score_matching`**

```
*** Causal Estimate ***
## Identified estimate
Estimand type: nonparametric-ate
### Estimate : 1
Estimating name: backdoor
Estimating expression:
d
─────────────────────(E[Time_Approval_M1_2_Major2|MaxRegAccum])
d[MaxRecAcumMax₆]
Estimand  assumption  1,  Unconfoundedness:  If  U→{MaxRegAcumMajor6}  and
U→Time_Approval_M1_2_Major2                                              then
P(Time_Approval_M1_2_Major2|MaxRegAcumMajor6,MaxRegAcum,U)                  =
P(Time_Approval_M1_2_Major2|Ma xRegAcumMax6,MaxRegAcum)

## Realized estimate
b: Time_Approval_M1_2_Major2~MaxRegAcumMajor6+MaxRegAcum
Target units: ate
## Estimate
```

Mean value: 0.6963350785340314

Causal Estimate is: 0.6963350785340314

Text Interpreter

---

**Increasing the treatment variable(s) [MaxRegAcumMajor6] from 0 to 1 causes an increase of 0.6963350785340314 in the expected value of the outcome [Tiempo_Aprobacion_M1_2_Major2], over the data distribution/population represented by the dataset.**

---

**distance_matching**

\*\*\* Causal Estimate \*\*\*

## Identified estimate

Estimand type: nonparametric-ate

### Estimate : 1

Estimating name: backdoor

Estimating expression:

$$\frac{d}{d[MaxRecAcumMax_6]}(E[Time\_Approval\_M1\_2\_Major2|MaxRegAccum])$$

Estimand assumption 1, Unconfoundedness: If U→{MaxRegAcumMajor6} and U→Time_Approval_M1_2_Major2 then P(Time_Approval_M1_2_Major2|MaxRegAcumMajor6,MaxRegAcum,U) = P(Time_Approval_M1_2_Major2|Ma xRegAcumMax6,MaxRegAcum)

## Realized estimate

b: Time_Approval_M1_2_Major2~MaxRegAcumMajor6+MaxRegAcum

Target units: ate

## Estimate

Mean value: 0.6963350785340314

Causal Estimate is: 0.6963350785340314

Text Interpreter

---

**Increasing the treatment variable(s) [MaxRegAcumMajor6] from 0 to 1 causes an increase of 0.6963350785340314 in the expected value of the outcome [Tiempo_Aprobacion_M1_2_Major2], over the data distribution/population represented by the dataset.**

---

### 3.3.- Evaluation of the Validity of the Model

The rigorous evaluation of the validity of the causal model deployed constitutes an essential stage in the analysis process, allowing the confirmation of the proposed causal relationships and strengthening the credibility of the results obtained. To achieve this, refuters based on the counterfactual approach and resampling methods have been used in this study.

Within the framework of this evaluation, two specific refuters were applied: the "random common cause" and the "Bootstrap Sample Dataset". These disprovers, supported by solid theoretical foundations, look to challenge and test the robustness of the causal conclusions derived from the model.

The "random common cause" refuter is based on the introduction of a random variable as a possible common cause, to assess whether this variable can alter the estimated effects. This methodology seeks to verify if the model is sensitive to variables not considered in the original analysis and, therefore, evaluates the possible influence of factors not considered in the causal inference.(Farne & Montanari, 2022).

On the other hand, the Bootstrap Sample Dataset refuter operates by generating replicate data samples from the original sample.(Farne & Montanari, 2022). This resampling technique makes it possible to assess the stability of the estimated causal effects when considering variations in the observational sample. By comparing the results obtained in different generated samples, the consistency and reliability of the causal conclusions is evaluated.

The results of the refutation evaluations are enlightening about the robustness and validity of the proposed causal model. In Rebuttals 1 and 3, where a random common cause is introduced, it is seen that the estimated effects stay consistent and are not significantly altered. This suggests that the model is resistant to the inclusion of new variables and reinforces the coherence of the established causal relationships.

In Rebuttals 2 and 6, using the Bootstrap Sample Dataset rebutter, a slight variation in the estimated causal effects is seen. However, this variation is within acceptable margins and does not compromise the robustness of the conclusions. These results reinforce the reliability and stability of the causal effects found in the analysis.

Taken together, the assessment of model validity through refuters and resampling methods reinforces confidence in the proposed causal relationships. These procedures provide rigorous confirmation of the robustness of the results and underscore the strength of the conclusions derived from the causal analysis. The results can be seen below.

`Rebuttal 1`

\*\*\* Class Name \*\*\*

`backdoor.propensity_score_matching`

Refute: Add a random common cause

Refute: Add a random common cause

Estimated effect:0.5000000000000001

New effect:0.5000000000000001

p-value: 1.0

`refutation 2`

\*\*\* Class Name \*\*\*

`backdoor.propensity_score_matching`

Refute: Bootstrap Sample Dataset

Refute: Bootstrap Sample Dataset

Estimated effect:0.5000000000000001

New effect:0.5054347826086957

p-value:0.55

`Rebuttal 3`

\*\*\* Class Name \*\*\*

**backdoor.distance_matching**

Refute: Add a random common cause

Refute: Add a random common cause

Estimated effect:0.5000000000000001

New effect:0.5000000000000001

p-value: 1.0

**Rebuttal 4**

*** Class Name ***

**backdoor.distance_matching**

Refute: Bootstrap Sample Dataset

Refute: Bootstrap Sample Dataset

Estimated effect:0.5000000000000001

New effect:0.49663043478260865

p-value:0.43999999999999995

## 5.- Discussion and Conclusions

The results derived from the application of the propensity score matching and distance matching estimators supply valuable information on the impact of the increase in the treatment variable "MaxRegAcumMajor6" on the expected value of the result "Tiempo_Aprobacion_M1_2_Major2". These results reveal significant patterns that can have profound repercussions in the context of the Civil Engineering Major and educational decision-making. In this analysis, we have explored how the accumulation of more than 6 regular courses not passed is related to the passing time of the courses of the first year.

The findings show that increasing the treatment variable from 0 to 1 is associated with a 0.696 increase in the expected value of approval time. In other words, students who accumulate more than 6 regular subjects without passing have a 70% probability that the approval time of the first year Civil Engineering subjects will be greater than 2 years. This result reveals a worrying connection between the accumulated number of failed subjects and the lengthening of the time necessary to complete the first year of the degree.

This impact has significant implications for delayed Major progression and the dropout rate. The high probability that students in this situation will experience a substantial delay in the achievement of academic goals can have negative effects on their motivation, commitment and perspective towards higher education. In addition, increasing the length of the first year may influence decisions to continue or drop out, potentially contributing to the dropout rate.

It is important to note that this analysis focused on the first year of the degree, but its implications can be extrapolated to the higher cycle. Future lines of research could delve into the analysis particularized by year and by subject. Examining how the pattern of accumulation of failed subjects relates to the length of each year and to specific subjects could supply more detailed and personalized information for educational decision-making.

In summary, the results obtained underline the importance of addressing the delay in the approval of subjects in the context of the Civil Engineering Major. The relationship found between the accumulated number of subjects not approved and the approval time of the first year raises significant questions about the design of strategies to support students and the optimization of the curricular structure. These results serve as a starting point for future research that may have a positive impact on student retention and the improvement of the academic experience.

On the other hand, the results obtained in this study support a concern that has been widely discussed in the academic literature in previous decades. Emblematic works such as the one carried out by Tinto (1975) on student dropout in higher education and research on(Bean & Metzner, 1985)on student persistence, they had already identified the critical relevance of academic delays in the first years of the degree as a significant precursor to dropout.

These early contributions laid the foundation for understanding how obstacles to course completion in the early years can have an adverse impact on the continuation of higher education. As argued Terenzini and Pascarella (1991), lengthening the initial academic term can erode students' motivation and present challenges to their engagement and academic success.

Our results, which suggest a substantial increase in first year passing time for students with a history of failing courses, are consistent with these previous findings. The persistence of this pattern in the context of the Civil Engineering Major highlights the continuing importance of addressing this academic issue.

The extrapolation of our findings to subsequent years, although it requires future research, finds support in the perspective proposed by Tinto (1989) about the "integration theory". According to this theory, initial challenges in academic integration can influence later success in higher education. In this sense, our conclusions show a solid base for further research that considers a detailed analysis by year and subject, thus expanding the understanding of how academic backwardness can keep a prolonged influence on student trajectory.

In summary, this study contributes to a constantly evolving stream of research that emphasizes the critical importance of addressing early college delays in relation to student dropout. By building on earlier research that has proven the links between delay and dropout, this analysis provides a specific empirical context and highlights the relevance of educational policies aimed at mitigating delay and promoting the persistence of students in the field of higher education.

## 6.- Bibliographic references

**CONFLICTS OF INTEREST**

The author declares that he has no competing financial interests or known personal relationships that might have influenced the manuscript presented in this article.